# On the "usual" misunderstandings between econophysics and finance: Some clarifications on modelling approaches and efficient market hypothesis

Marcel Ausloos, Franck Jovanovic, Christophe Schinckus
School of Management – University of Leicester

## Abstract

In line with the recent research and debates about econophysics and financial economics, this article discusses on usual misunderstandings between the two disciplines in terms of modelling and basic hypotheses. In the literature devoted to econophysics, the methodology used by financial economists is frequently considered as a top-down approach (starting from a priori "first principles") while econophysicists rather present themselves as scholars working with a (empirical data prone) bottom-up approach. Although this dualist perspective is very common in the econophysics literature, this paper claims that the distinction is very confusing and does not permit to reveal the essence of the differences between finance and econophysics. The distinction between these two fields is mainly investigated here through the lens of the Efficient Market Hypothesis in order to show that, in substance, econophysics and financial economics tend to have a similar approach implying that the misunderstanding between these two fields at the modelling level can therefore be overstepped.

## I) Introduction

This article is a contribution in line with the recent research aiming to increase the dialogue between physicists (particularly econophysicists) and financial economists (Ausloos, 2001, 2013; Bouchaud, 2002; Bouchaud et al., 2002; Carbone et al., 2007; Chakrabarti & Chakraborti, 2010; Chen & Li, 2012; Durlauf, 2005, 2012; Farmer & Lux, 2008; Gabaix, 2009; Jovanovic & Schinckus, 2015, 2016; Keen, 2003; Lux, 2009; McCauley, 2006, 2009; McCauley et al., 2007; Potters & Bouchaud, 2003; Sornette, 2014; Stanley & Plerou, 2001). Actually, a recent article by Sornette (2014) offers a titillating example of largely widespread confusions about the distinction between econophysics and financial economics. We claim here that the cross-fertilization between econophysics and financial economics requires an objective clarification of both approaches in order to open the door for an interdisciplinary and fruitful dialogue.

The misunderstanding evoked above seems to be rooted in the difference, pointed out by Sornette (2014), between the way of modelling in economics and how it is done in physics, which is broadly resumed by the "*difference between empirically founded science and normative science*" (Sornette, 2014, p. 3). As explained,

> "The difference between [the model for the best estimate for the fundamental price from physics] and [the model for the best estimate for the fundamental price from financial economic, i.e. efficient market theory] is at the core of the difference in the modeling strategies of economists, that can be called top-down (or from rational expectations and efficient markets), compared with the bottom-up or microscopic approach of physicists" (Sornette, 2014, p. 7).

This distinction between the ways of modelling provides the corner argument for explaining the major differences between the two disciplines. Actually, this opposition is also used for claiming that modelling in economics can be looked on as a "puzzle" which

> "refers to problems posed by empirical observations that do not conform to the predictions based on theory" (Sornette, 2014, p. 5).

In order to give up such a kind of puzzle, Sornette (2014, p. 7) suggested to use this distinction for formulating an (econo)physics definition of the efficient market hypothesis (EMH) compatible with the bottom-up approach. We thoroughly re-examine this "solution" from a fundamental conceptual way and from a critical analysis of the EMH.

This dualist perspective (top-down vs. bottom-up) is frequently found in the econophysics literature (Bouchaud & Challet, 2014; Bouchaud & Potters, 2003; Challet et al., 2005, p. 14; McCauley, 2004, 2006; Rickles, 2008; Schinckus, 2010; Stanley et al., 1999, p. 157). It is also a common argument for questioning the use of the Gaussian framework in a large number of financial economists' works. However, the argument based on this distinction is very confusing. It could even shock some financial economists who face regularly critiques about the too micro-focused knowledge usually implemented in finance. In the same vein, from a financial point of view, a lot of publications from econophysicists appear to be mainly phenomenological focusing on a macro-description of financial markets/economic systems.

This classical opposition between econophysics and financial economics cannot be reduced to a couple of dualism: empirical science vs. normative science or micro vs. macro perspectives. Actually the opposition between these two fields is not obvious. We will investigate this point by arguing that, surprisingly, econophysicists and financial economists use a quite similar approach for analysing financial markets. However, their approaches appear to be different because these scientists are trained in "different schools" with different aims/vocabularies. We will illustrate this aspect in the following sections, yet wondering if it is only a matter of words. We will conclude by claiming that econophysics and financial economics tend to have a similar approach, although their respective backgrounds lead to present it differently. Therefore, the misunderstanding between these two fields can be overstepped.

**II) Is it only a matter of words?**

No. This scientifically antagonistic situation is not only a matter of words. The gap between econophysicists and financial economists is deeper than just a problem of translation. We rather demonstrate that the misunderstanding so often evoked in the literature, results from an inappropriate comparison between the methodology (i.e. use of assumptions) and the modelling process (i.e. how to implement assumptions/approach) used by economists and physicists. According to some econophysicists, the "modelling strategies of economists can be called top-down" because "financial modeler builds a model or a class of models based on a pillar of standard economic thinking, such as efficient markets, rational expectations, representative agents" (Sornette, 2014, p. 5). However, it is worth mentioning that rational expectations or efficient market are hypotheses about the economic reality. In accordance with this formulation, economists usually start their analysis with assumptions that they implement in their way of modelling and from which they deduce conclusions that are tested. In other words, Sornette associated what he called "top-down" modelling with the hypothetico-deductive approach mainly used in economics.

Regarding this methodological aspect, econophysics is, in contrast, often presented as a data-driven field founded on descriptive models resulting from observations. First of all, it is important to remind that the belief in *no a priori* is, on itself, an a priori since it refers to a form of positivism[1]. The positivism (research methodology) takes the stance that the scientist knowledge exists independently from the social actors. In opposition to these, the interpretivism is more concerned about the different views that (social) agents have towards some phenomenon. Secondly, the inductive and data driven approach implemented by econophysicists can be seen as a "bottom up" methodology because it starts with data related to a specific phenomenon rather than starting with assumptions on it. The recurrence of observations allows scholars to make some generalizations for other similar phenomena. Why does it matter to mention this point? Simply to clarify that the words "top down" and "bottom up" refer to methodological strategies and not to a way of modelling.

The difference between methodology and modelling seems subtle, but it is important: the methodology refers to the conceptual way of dealing with phenomena, i.e. quantitatively vs. qualitatively; empirically vs. theoretically. By contrary, the modelling way rather concerns the kind of computation (and data) used by scientists. A opposition between econophysics and financial economics with a confusion on these two levels led several authors (McCauley, 2006; Sornette, 2014) to deal with a very specific part of the literature related to these two areas of knowledge. On the one hand, although a part of economics is well-known for its representative agent modelling, this field does not necessary implement a top-down modelling as claimed by Sornette. Agent based modelling, for instance, is a common practice in economics and financial economics (Chen, 2012; Gilbert, 2007; LeBaron, 2000, 2006; Tesfatsion, 2001, 2003), enforcing the micro-oriented modelling already used in these fields. On the other hand, a large part of the literature in econophysics is dedicated to the phenomenological macro-description of the evolution of financial prices making these works inappropriate for being considered as "bottom up" modelling. One can mention, among other works, Takayasu and Takayasu (2016) who observed that the large fluctuations on the financial markets can be captured through a power law, while Levy (2003) or Klass et al (2006) confirmed the conclusion made by Pareto (1897) more than one century ago showing that wealth and income distribution can both statistically be characterized by a power law[2]. In the same vein, Amaral et al. (1997) explained that the annual growth rates for US manufacturing companies can also be described through a power law whereas Axtell (2001), Luttmer (2007) or Gabaix and Landier (2008) claimed that this statistical framework can also be used to characterize the evolution of the firms size as a variable of their assets, market capitalization or number of employee. These "size models" have afterwards been applied for describing the evolution of the cities size (Cordoba, 2008; Eeckhout, 2004; Gabaix, 1999). Although this phenomenological tradition considers that economic systems are composed of multiply interacting components (no learning agents), these components are assumed to interact in such a way that they generate macro-properties for systems (Rickles, 2008). Thus, because this methodology induces macro-properties in terms of statistical regularities without defining in details all aspects of the micro-level, it cannot really be considered as a bottom-up approach.

---

[1] Different views about "knowledge" exist of course, for further information about this point, see (Saunders et al., 2012, p. chap. 4).

By misunderstanding the epistemic role of the key concepts in financial economics, such as the EMH, rational expectations etc., some econophysicists criticize financial economics and emphasize its weaknesses from a perspective that is not at all an issue for financial economists. Moreover, as Lux (2009, p. 230) already recalled

> "[o]ne often finds [in the literature from econophysics] a scolding of the carefully maintained straw man image of traditional finance. In particular, ignoring decades of work in dozens of finance journals, it is often claimed that "economists believe that the probability distribution of stock returns is a Gaussian", claim that can easily be refuted by a random consultation of any of the learned journals of this field".

True! It is well known that economists identified stylized facts on stock price extreme variations and their leptokurticity several decades before the emergence of econophysics (Bowley, 1933; Houthakker, 1961; Larson, 1960; Mills, 1927; Mitchell, 1915; Olivier, 1926)[3]. In the same vein, since the 1970s, financial economists have developed several models to take into account extreme values, like jump-diffusion models (Merton, 1976; Press, 1967) or the ARCH types models empirically implemented by financial economists (Bollerslev, 1986; Engle, 1982)[4].

In other words, we claim that the dualism top-down vs. bottom-up is not appropriate to oppose financial economics and econophysics because this criterion is usually applied at two different levels: when econophysics consider that financial economics is based on a top-down approach, they usually refer to the methodology used by economists; – in contrast, when they claim they are using a bottom-up approach, econophysicists are arguing on their way of modelling as in statistical physics. This confusion can therefore lead to a deaf dialogue and an intellectual conservatism narrowing the possibility of interactions between both communities.

---

[2] We can also mention Mantegna and Stanley (1994), Lux (1996), Bak et al. (1997), Ausloos (2000), Gligor and Ignat (2001), Alvarez-Ramirez et al. (2001), Alvarez-Ramirez et al. (2002), Hsu and Lin (2002), Gabaix et al. (2003), and Yura et al. (2014).
[3] See Lux (2009) or Jovanovic and Schinckus (2013) for further details.
[4] See Francq and Zakoian (2010), Bauwens et al. (2006), Tim (2010) and Pagan (1996) for further details on these categories of models.

**III) Comparing apples with pears**

Sornette rooted the opposition between econophysics and financial economics in the three inter-related pillars of contemporary physics that are "experiments, theory and numerical simulations" (Sornette, 2014, p. 2). It is worth mentioning that these methodological "pillars" also exist in economics. Experimental economics is a very well know field in economics (Guala, 2008; Roth, 1993; Smith, 1992). We can also mention that in 2002 Smith and Kahneman received the 'Nobel prize' distinction for their contributions in experimental economics. Moreover, experimental finance has its own journal, the *Journal of Behavioral and Experimental Finance.* Computational economics (including computational financial economics) is also a well know active field (Amman et al., 1996; Bloomfield, 2010; Miranda & Fackler, 2002) with its own journals like the *Computational Economics* (the journal of the Society for Computational Economics) or the *Journal of Economic Dynamics and Control*. However, by making wrongfully comparisons between two different conceptual levels, many econophysicists fuel the misunderstanding: indeed, it is very common for econophysicists to compare the results of their statistical models with the theoretical framework of financial economics (efficient market hypothesis). For instance, when Sornette explains that econophysics deals with a microscopic approach, he implicitly discusses the way of dealing with data (experimental level) that he compared with the EMH which refers to the theoretical level and not to an experimental level. In this context, it is a truism to claim that the experimental level is closer to reality than theoretical one (whatever the field). While we acknowledge that these three conceptual levels are interrelated, an interdisciplinary comparison makes sense only if one compares the same conceptual levels.

If econophysicists wish to compare (the statistical) models used in both disciplines, they should compare their models with ARCH class of models. Surprisingly, even though econophysicists and economists do not have in general the same methodology (the former are data-driven while the latter start with assumptions), both communities of researchers do proceed in a similar way regarding the implementation of models: both follow a bottom-up approach by calibrating their models in order to simulate features (e.g., price or return variations). The difference refers to the way of modelling the extreme values: econophysicists consider these values as a part of the system while economists rather associate these extreme values with an error terms to which they give a statistical distribution. In other words, economists mainly describe stylized facts in two steps: 1) a general trend (assumed to be ruled by a Brownian uncertainty) whose 2) variations follow a conditional distribution for which a calibration is required. For these conditional distributions, it goes without discussion that data-driven calibration (i.e. without theoretical justification) is common in financial economics. In fact, such an approach has also generated methodological debates among economists, for instance, about the relevance of a strictly data-driven approach in the field: one can mention the Koopmans-Vining debate at the end the 1940s (Mirowski, 1989), or more recently, the Vector Autoregressions (VARs) modelling (Chari et al., 2008, 2009; Christiano, 2012) or the Real Business Cycle (RBC) models approach (Eichenbaum, 1996; Hansen & Heckman, 1996; Hoover, 1995; Quah, 1995; Sims, 1996). The ARCH class of models, which is a statistical modelling approach, based on unlimited arbitrary inputs without theoretical interpretations, has faced with the same criticisms as that of physicists, from a financial economists' viewpoint (Pagan, 1996), particularly because these models can "not provide an avenue towards an *explanation* of the empirical regularities" (Lux, 2006).

Nevertheless, although both communities use a similar calibration approach, one can observe a paradoxical situation: financial economists accept calibration models (as ARCH class modeling for instance) but they have difficulties to accept it for the models coming from statistical physics that are more based on physics conceptual ideas than statistical considerations. This position needs to be clarified. Jovanovic and Schinckus (2015, 2016) have detailed and explained this paradox by showing that the major differences between the two fields appears in the way of calibrating their models. Roughly speaking, calibration made in econophysics mainly results from data whereas it is rather founded on a statistical assumption (Brownian uncertainty) in finance. Concretely, ARCH models are closely combined with a theoretical explanation since they were introduced in finance with the purpose to test the EMH. This hypothesis constitutes one of the major theoretical foundations of the financial economics' framework (Fama, 1991; Jovanovic, 2002, 2010; Jovanovic & Schinckus, 2016; Malkiel, 1992). It is worth mentioning that the EMH is hardly testable and that any empirical test of this hypothesis refers to what it is called in the literature a joint-test. A joint-test refers to the fact that, on a given market, any test of the efficiency (i.e. the fact prices fully reflect available information) tests at the same time the notion of efficiency and the asset-pricing model used to price securities on this market. In other words, any empirical refutation can be due to either the fact that the market is not efficient or the model used is not appropriate for the test[5].

Such a joint-test implies that market efficiency per se is not testable (Campbell et al., 1997; Cuthbertson, 2004; Fama, 1976; Jovanovic, 2010; LeRoy, 1976, 1989; Lo, 2000), and any test of efficiency has to be considered carefully. We will clarify this point afterwards. This link between ARCH models and the EMH explains why most of financial economists using ARCH class of models consider their models have theoretical foundations from a financial (and not from statistical only) point of view. In this perspective, calibration in financial economics appears as an empirical test of theoretical hypothesis.

At this stage, econophysicists and financial economists have the same procedure for modelling (calibration of models to fit data) but the latter combine the calibration step

---

[5] We can mention that this joint-test is similar to the question arising from Benford law application: is a lack of conformity either intrinsic to the analysed system or is it because Benford should not apply? (Ausloos et al., 2014; Raimi, 1976)

with a specific theoretical framework while the former claim to be more empirical data-driven. By combining a theoretical framework to set up the initial calibration of the formalized systems, the "model becomes an a priori hypothesis about real phenomena" (Haavelmo, 1944, p. 8). Although some econophysicists (McCauley, 2006; Sornette, 2014) criticize this theoretical dependence of the modelling in finance, it is worth mentioning that physics also provides telling similar examples in which a theoretical framework is accepted while the empirical results are wholly incompatible with this framework. One could mention the example of the more recent Higgs Boson discovery (Allen, 2014; Allen & Lidström, 2015). The concept of the Higgs Boson pre-existed to its observation meaning that its theoretical framework was assumed during several years without observing this particle (Morrison, 2015). In the same vein, the often discussed string theory is an elegant mathematical framework whose empirical/concrete evidences are still on debates (Aganagic, 2016). This is not the unique counterexample, "there are plenty of physicists who appear to be unperturbed about working in a manner detached from experiment: quantum gravity, for example. Here, the characteristic scales are utterly inaccessible, there is no experimental basis, and yet the problem occupies the finest minds in physics" (Rickles, 2008, p. 14).

**IV) What is wrong with EMH?**

After having nuanced the (not so different) ways of modelling in the two fields, we can now examine the common critique made by econophysicists about the EMH. Actually, Sornette's paper (2014) is also a telling example of the critiques econophysicists address to EMH showing another confusion relative to the status of this hypothesis. Broadly, why keeping the EMH given it is refuted by the fact that stock market variations are not Gaussian. In this perspective, the EMH is often presented as an a priori hypothesis by econophysicists. However, the status of this hypothesis is confusing. The EMH has a very specific place in financial economics that is unclear for most of econophysicists as well as for most of financial economists, particularly because the EMH is generally identified to the Brownian motion or to the random character of stock market variations.

The EMH was proposed in the 1960s on the intuition that a pure random-walk model would verify two properties of competitive economic equilibrium: the absence of marginal profit and the equalisation of a stock's price and value, meaning that the price perfectly reflects the available information[6]. This project was undeniably a *tour de force* at that time: creating a hypothesis that made it possible to incorporate econometric results and statistics on the random nature of stock-market variations into the theory of economic equilibrium (Jovanovic, 2008, 2010; Sewell, 2011). It is through this link that one of the main foundations of current financial economics was laid down and that the importance of the pure random-walk model, or Brownian motion, and thus of the Gaussian distribution, can be explained: validating the random nature of stock-market variations would, in effect, establish that prices on competitive financial markets are in permanent equilibrium as a result of the effects of competition. This is what the EMH should be: the random character of stock market variations would imply that the prices reflect the competitive equilibrium by incorporating the available information.

---

[6] It is worth mentioning that while use of a random-walk model to represent stock-market variations was first proposed in 1863 by a French stockbroker, Jules Regnault, and then formalized by the mathematician Louis Bachelier (Bachelier, 1900; Davis & Etheridge, 2006), the EMH was created in the 1960s.

Unfortunately, this hypothesis does not really reach this goal. To establish the link between the empirical observations on the random character of stock-market variations, a stochastic model (i.e. Brownian motion) and the theory of economic equilibrium, Fama extended the early theoretical thinking of the 1960s and transposed onto financial markets the concept of free competitive equilibrium on which rational agents would act (1965b, p. 56). Such a market would be characterised by the equalisation of stock prices with their equilibrium value. This value is determined by a valuation model; the choice of this model is irrelevant for the EMH. In Fama's thesis, this equilibrium value is the fundamental – or intrinsic – value of a security. The signification of this value is unimportant: it may be the equilibrium value determined by a general equilibrium model, or a convention shared by "sophisticated traders" (Fama, 1965a, pp. 36, fn 33).

Notwithstanding, Fama later dropped the reference to a convention and stated that the equilibrium model valued stocks using all available information in accordance with the idea of competitive markets. Following the most commonly accepted definition of efficiency proposed by Fama in his 1970 paper, on an efficient market, equalisation of the price with the equilibrium value meant that any available information was included in prices. Consequently, that information has no value in predicting future price changes: future prices are independent of past prices. For this reason, Fama considered that, in an efficient market, price variations should be random, like the arrival of new information, and that therefore implies an impossibility to beat the market (Fama, 1965a, p. 35 or 98). A random-walk model made possible to simulate dynamic evolution of prices in a free competitive market that is in constant equilibrium. In other terms, the market is Markovian; it has no memory.

Fama started his demonstration from empirical observations, particularly the existence of different agents and behaviours on financial markets. According to him, "there is no strong reason to expect that each individual's estimates of intrinsic values will be independent of the estimates made by others (i.e., noise may be generated in a dependent fashion). For example, certain individuals or institutions may be opinion leaders in the market. That is, their actions may induce people to change their opinions concerning the prospects of a given company" (Fama, 1965a, p. 37). Consequently, for the purpose of demonstrating these properties, Fama assumed the existence of two kinds of traders: the "sophisticated traders" and the "others". In this perspective, "sophisticated traders" are those who influence the market. Fama's key assumption was that "sophisticated traders", due to their skills, make a better estimate of the intrinsic/fundamental value than other agents do by using all available information. Moreover, Fama assumes that, "although there are sometimes discrepancies between actual prices and intrinsic values, sophisticated traders in general feel that actual prices usually tend to move toward intrinsic values" (1965a, p. 38). Since "sophisticated traders" share the same valuation model for asset prices (Fama, 1965a, p. 40), their transactions will help prices trend towards the fundamental value. Fama added, using arbitrage reasoning, that any new information is immediately reflected in prices (1965a, p. 39). The independence of price variations, resulting from the random arrival of new information, and the absence of profit being two characteristics of the random walk model allow Fama to make a direct connection between this model and the market efficiency. In other words, by assuming that sophisticated traders' have financial abilities superior to those of other agents, Fama showed that the random nature of stock-market variations is synonymous with dynamic economic equilibrium in a free competitive market.

But when the time came to demonstrate mathematically the intuition of the link between information and the random (independent) nature of stock-market variations, Fama became elusive. He explicitly attempted to link the EMH with the random nature of stock-market variations in an article published in 1970. Seeking to generalise, he dropped all direct references to the notion of "fundamental value" and to "sophisticated traders". Consequently, all agents were assumed to use the same model for evaluating the price of financial assets (i.e. representative agent hypothesis). Finally, he kept the general hypothesis that "the conditions of market equilibrium can (somehow) be stated in terms of expected returns" (1970, p. 384). He formalised this hypothesis by using the definition of a martingale:

$$E\left(\tilde{p}_{j,t+1}\vert\Phi_t\right) = \left[1 + E\left(\tilde{r}_{j,t+1}\vert\Phi_t\right)\right]p_{j,t}, \text{ with } r_{j,t+1} = \frac{p_{j,t+1} - p_{j,t}}{p_{j,t}}, \tag{1}$$

where the tilde indicates that the variable is random, $p_j$ and $r_j$ represent the price and return for a period of the asset $j$, $E(./.)$ the conditional expectation operator, and $\Phi_t$ represents all information at the time $t$.

This equation implies that "the information $\Phi_t$ would be determined from the particular expected return theory at hand" (1970, p. 384). Fama added that "this is the sense in which $\Phi_t$ is 'fully reflected' in the formation of the price $p_{j,t}$" (1970, p. 384). To test the hypothesis of information on efficiency, he suggested that from this equation one can obtain the mathematical expression of a fair game, which is one of the characteristics of a martingale model and a random-walk model. A demonstration of this link would ensure that a martingale model or a random-walk model could test the double characteristic of efficiency: total incorporation of information into prices and the nullity of expected return.

This is the most well-known and used formulation of the EMH. However, it is important to mention that the history of the EMH went beyond the Fama (1970) article. Indeed, in 1976, LeRoy showed that Fama's demonstration is tautological and that his hypothesis is not testable. Fama answered by changing his definition and admitted that any test of the EMH is a test of both market efficiency and the model of equilibrium used by investors (Fama, 1976). Moreover, he modified his mathematical formulation:

$$E_m\left(\tilde{R}_{j,t}\vert\Phi_{t-1}^m\right) = E\left(\tilde{R}_{j,t}\vert\Phi_{t-1}\right) \tag{2}$$

where $E_m(\tilde{R}_{j,t}|\Phi_{t-1}^m)$ is the equilibrium expected return on security *j* implied by the set of information now used by the market at *t* – 1, $\Phi_{t-1}^m$, and $E(\tilde{R}_{j,t}|\Phi_{t-1})$ is the true expected return implied by the set of information available at *t* – 1, $\Phi_{t-1}$. From then on, efficiency presupposed that, using Fama's own terms, the market "correctly" evaluates the "true" density function conditional on all available information. Thus, in an efficient market, the truly perfect model for valuing the equilibrium price is available to agents. To test efficiency, Fama reformulated the notion of "expected return" by introducing a distinction between price – defined by the true valuation model – and agents' expectations. The test consisted in verifying whether the return expected by the market based on the information used, $\Phi_{t-1}^m$, is equal to the expectation of true return obtained on the basis of all information available, $\Phi_{t-1}$. This true return is obtained by using the "true" model for determining the equilibrium price.

Fama proposed testing the efficiency in two ways, both of which relied on the same process. The first test consisted in verifying whether "trading rules with abnormal expected returns do not exist" (1976, p. 144). In other words, this was a matter of checking that one could obtain the same return as that provided by the true model of assessment of the equilibrium value on the one hand and the set of available information on the other hand. The second test would look more closely at the set of information. It was to verify that "there is no way to use the information $\Phi_{t-1}$ available at *t*-1 as the basis of a correct assessment of the expected return on security *j* which is other than its equilibrium expected value" (1976, p. 145).

At the close of his 1976 article, Fama answered LeRoy's criticisms: the new definition of efficiency was *a priori* testable (we will precise this point hereafter). It should be noted however that the definition of efficiency had changed: it now referred to the true model for assessing the equilibrium value. For this reason, testing efficiency required also testing that agents were using the true assessment model for the equilibrium value of assets. Fama acknowledged the difficulties involved in this joint test in a report on efficiency published in 1991 (Fama, 1991, pp. 1575-1576). The test would, then, consist in using a model for setting the equilibrium value of assets – the simplest would be to take the model actually used by operators – and determining the returns that the available information would generate; then to use the same model with the information that agents use. If the same result is obtained – that is, if equation (2) is verified – then all the other information would indeed have been incorporated into prices. It is striking to note that this test is independent of the random nature of stock-market variations. This is because, in this 1976 article, there is no more talk of random walk or martingale; no connection with a random process is necessary to test efficiency.

Despite of this important conclusion, Fama's article (1976) is practically not cited. Almost all financial economists refer to the 1970 article and keep the idea that to validate the random nature of stock-market variations means to validate market efficiency, fuelling the confusion between the market efficiency and the random character of stock market variations. However, this problem was rapidly pointed out. LeRoy (1973) and Lucas (1978) provided theoretical proofs that efficient markets and the martingale hypothesis are two distinct ideas: a martingale is neither necessary nor sufficient for an efficient market. In a similar way, Samuelson (1973), who gave a mathematical proof that prices may be permanently equal to the intrinsic value and fluctuate randomly, explained that the making of profits by some agents cannot be ruled out, contrary to the original definition of the EMH. In the same vein, De Meyer and Saley (2003) showed that stock-market prices can follow a martingale even if all available information is not reflected in the prices. We can also mention Cutler, Poterba and Summers (1989), Longin (1996) and Cornell (2013) who showed that large daily price movements are not always related to the arrival of new information. In other terms, the EMH must be clearly dissociated from stochastic processes, including the Gaussian ones.

Unfortunately, a proliferation of theoretical developments combined with the accumulation of the empirical works led to a confusing situation. Indeed, the definition of efficient markets has changed depending on the emphasis placed by each author on a particular feature. For instance, Fama *et al.* (1969, p. 1) defined an efficient market as "a market that adjusts rapidly to new information"; but Jensen (1978, p. 96) considered that "a market is efficient with respect to information set $\theta_t$ if it is impossible to make economic profit by trading on the basis of information set $\theta_t$"; while according to Malkiel (1992) "the market is said to be efficient with respect to some information set […] if security prices would be unaffected by revealing that information to all participants. Moreover, efficiency with respect to an information set […] implies that it is impossible to make economic profits by trading on the basis of [that information set]".

To sum up, the EMH is a theoretical assumption that aims at giving a theoretical meaning to the random character of stock markets observed in the 1960s and at creating a scientific framework for finance (i.e. the financial economics) (Findlay & Williams, 2001; Jovanovic, 2008). However, it is important to keep in mind that the EMH and the random character of stock markets are two different elements: the Gaussian dimension of data is neither a necessary nor a sufficient condition for having EMH.

By keeping the confusion between the EMH and the Gaussian stochastic processes authors, like Sornette (2014, section 3.1), believe that the evidence about the non-Gaussian distribution of financial data should lead to reject the EMH. While Gaussian stochastic processes are not a valid test of the EMH, such a position misses in addition the fact that "prices always 'fully reflect' available information" (Fama, 1970, p. 383) is an ideal that does not exist. Precisely, as the next section will precise, EMH is rooted into the positivism defended by Friedman (Findlay & Williams, 2001)[7]. Given the contradictions discussed previously, econophysicists (and also financial economists) could legitimately ask why financial economists keep the EMH and the Gaussian stochastic processes and discuss them at length through papers and doctoral theses.

---

[7] Even theoretically, EMH is an ideal that does not exist. Indeed, very soon economists demonstrated that the mechanism of the market efficiency contains a theoretical contradiction and consequently, informationally efficient markets are impossible (Grossman, 1976; Grossman & Stiglitz, 1976, 1980).

## V) Can one give up EMH?

Roughly speaking, one can mention two major reasons explaining why financial economists keep EMH and its association with Gaussian process: 1) the role of the EMH in the construction of a social reality; 2) the use of statistical tests for validating a hypothesis or a model in economics.

The reason explaining why financial economists keep the EMH as a paradigm stems from a classical opposition between social sciences and physics (Eto, 2008). Whereas physicists are aware about their influence on the measure of physical phenomena, they do not influence the essence of the physical phenomena they study (i.e. the way the phenomena behave). Financial economics is a bit different since the financial reality is a social construction, which is built from conceptual frameworks. Consequently, scientists influence the way the phenomena they study behave. In this context, the EMH is an idealistic framework that has been used as a theoretical framework for implementing the computerization of financial markets (Schinckus, 2008), the international standardization of accounting conventions (Chane-Alune, 2006; Miburn, 2008), the legal policies in US (Hammer & Groeber, 2007; Jovanovic et al., 2015) and financial regulation policies (Muniesa, 2003; Pardo-Guerra, 2015).

In this perspective, coming back with our comments about methodology, models and concepts in finance have more interpretative role than the ones used in physics. Derman (2001, 2009) also emphasizes this opposition in his comparison of the way of modelling in finance and in physics: while physicists implement causal (drawing causal inference) or phenomenological (pragmatic analogies) models in their description of the physical world, financial economists use interpretative models to "transform intuitive linear quantities into non-linear stable values" (Derman, 2009, p. 30). Finances try "to solve a relative-value problem rather than an absolute-value problem" as in physics (Derman, 2001, p. 477). Physicists aim at describing a given world (i.e. a world that cannot be influenced by the observers) whereas economists try to interpolate a reactive system by pricing assets by using relative relations between the phenomena studied.

Using the EMH as an idealistic framework leads to conserve Fama (1970) definition in order to find ways for increasing the access to information or its incorporation into the prices. By contrary, using the definition suggested by Jensen (1978) or Malkiel (1992) which focuses on the possibility to make out-profit leads to increase the dialogue between econophysicists and financial economists. Precisely, the contemporary theoretical framework in finance (Harrison & Kreps, 1979; Harrison & Pliska, 1981) is not based on the definition of the EMH provided by Fama but rather on the absence of profit opportunity (called arbitrage-free) and of martingale measure. In this perspective, an arbitrage opportunity is a self-financing trading strategy such as a portfolio has a value equal to zero at the beginning can have a positive value at the end. Precisely, Harrison, Kreps and Pliska showed that a market is arbitrage-free (efficient) if there exists at least one martingale measure. This means that in a market free of arbitrage the stochastic price process for financial assets must have at least one martingale measure Consequently, keeping the clear distinction between the EMH and the Gaussian processes leads to maintain financial economics and econophysics in a common framework, inviting econophysicists to focus on the compatibility between their models and the contemporary theoretical framework in finance such as defined by Harrison, Kreps and Pliska (Harrison & Kreps, 1979; Harrison & Pliska, 1981).

The reason for keeping Gaussian stochastic processes refers to statistical tests. Statistical tests are a major scientific criterion for economists. To date, although econophysicists have developed several models mainly based on power laws, they traditionally use visual tests based on a double logarithmic axes histogram. These tests consist of comparing graphs deduced from observation with graphs deduced from the results of the models[8]. One of the reasons of this situation is that until the very recent period, no statistical test comparable to those used in financial economics was developed in the stable Levy framework, which implied to use the General Central Limit Theorem. Therefore, such tests are in their infancy, while Gaussian framework offers this opportunity for a very long time because they are based on the Central Limit Theorem.

Both approaches have their own drawbacks: financial economists are more concern by statistical tests, while econophysicists, originally trained in statistical physics, are more concerned by simulations as close as possible to empirical observations. In their perspective, econophysicists have mainly focused on visual comparisons, which generates significant systematic errors when we have to insure that they identify the considered variable distribution (Clauset, et al., 2009; Gillespie, 2014; Stumpf & Porter, 2012) and which do not have enough scientific foundations from the perspective of financial economists (Durlauf, 2005; Jovanovic & Schinckus, 2015, 2016; LeBaron, 2001).

While visual tests are the most common in the econophysics' literature, it is worth mentioning that some econophysicists use statistical tests. One can mention, among other works, Redelico et al. (2009) and Gligor and Ausloos (2007) who used the Student's t-test; Clippe and Ausloos (2012) and Mir et al. (2014) who used a chi-square test; and also Queiros (2005), Zanin et al. (2012), Theiler et al. (1992) or Morales et al. (2013). However, as some econophysicists pointed out, "better and more careful testing is needed, and that too much of data analysis in this area relies on visual inspection alone" (Farmer & Geanakoplos, 2008, p. 24).

Developing new statistical tests for non-Gaussian models is one of the major issues for the two disciplines. From a financial economics' viewpoint, several problems exist in order to develop statistical tests dedicated to power laws (Broda et al., 2013, p. 293). In this perspective, financial economists look dependant on their current models based on Gaussian distribution because of the tests currently available. The desire for developing new statistical tests is finally very poor, particularly because it could take time before publishing a good article in a top journal.

---

[8] For a discussion, see LeBaron (2001), Stanley and Plerou (2001), Mitzenmacher (2004), Newman (2005), Durlauf (2005), Clauset et al. (2009) and Jovanovic and Schinckus (2016).

## VI) CONCLUSION

This article has studied confusions that are largely widespread in the econophysics literature about financial economics, and vice versa. As we saw, both communities do not share the same scientific culture; this situation generated many oppositions between econophysicists and economists. These misunderstandings fuel the current deaf dialogue between the two scholar communities. However, this paper has shown the necessity to compare what is comparable. With this purpose, surprisingly, econophysics and financial economics have more in common than it is generally suggested in the literature. Part of their researches shares a calibration approach rooted in a bottom-up approach although their approaches are different. Concerning the EMH, by making a clear distinction between this hypothesis and the Gaussian processes the two disciplines can easily focus on common goals. For instance, while the rejection of Gaussian processes by econophysicists has led some of them (Lux & Ausloos, 2002; Vandewalle & Ausloos, 1997) to develop models away from the classical Brownian motion, going to fractional Brownian motion (and multifractals), we could expect that they try to integrate their results in the theoretical framework in finance such as defined by Harrison, Kreps and Pliska (Harrison & Kreps, 1979; Harrison & Pliska, 1981). As pointed out, financial economists are aware for keeping a link with the Gaussian framework in order to use statistical tests considered as strong enough. However, common research is still necessary in order to reduce the gap between the two frameworks and to develop fruitful collaborations.